\documentclass[useAMS,usenatbib]{mn2e}
\usepackage{longtable}
\usepackage{rotate}
\usepackage{lscape}
\usepackage{natbib}
\usepackage{latexsym,amssymb,verbatim}
\usepackage[pdftex]{graphicx}
\title[Deuterium chemistry of dense gas in the vicinity of low-mass and massive star forming 
regions]{Deuterium chemistry of dense gas in the vicinity of low-mass and massive star forming regions}
\author[Zainab Awad, Serena Viti, Estelle Bayet, and Paola Caselli]{Zainab Awad$^{1,2}$\thanks{E-mail:
zma@sci.cu.edu.eg} Serena Viti$^{2}$, Estelle Bayet$^{3}$, and Paola Caselli$^{4}$\\
$^{1}$Department of Astronomy, Space Science, and Meteorology, Faculty of Science, Cairo University, 
Giza 11326, Egypt\\
$^{2}$Department of Physics and Astronomy, University College London, London WC1E 6BT, UK\\
$^{3}$Sub-Department of Astrophysics, University of Oxford, Denys Wilkinson Building, Keble Road, 
Oxford, OX1 3RH\\
$^{4}$School of Physics and Astronomy, University of Leeds, Leeds LS2 9JT, UK}

\begin{document}
\date{Accepted 2014 June 9.  Received 2014 June 7; in original form 2013 October 8}
\pagerange{\pageref{firstpage}--\pageref{lastpage}} \pubyear{2014}
\maketitle 
\label{firstpage}
\begin{abstract}
The standard interstellar ratio of deuterium to hydrogen (D/H) atoms is $\sim 1.5 \times 10^{-5}$. 
However, the deuterium fractionation is in fact found to be enhanced, to different degrees, 
in cold, dark cores, hot cores around massive star forming regions, lukewarm cores, and 
warm cores ({\it hereafter}, hot corinos) around low-mass star forming regions. In this paper, we 
investigate the overall differences in the deuterium chemistry between hot cores and 
hot corinos. We have modelled the chemistry of dense gas around low-mass and massive 
star forming regions using a gas-grain chemical model. 
We investigate the influence of varying the core density, the depletion efficiency 
of gaseous species on to dust grains, the collapse mode and the final mass of the 
protostar on the chemical evolution of star forming regions. 
We find that the deuterium chemistry is, in general, most sensitive to variations of 
the depletion efficiency on to grain surfaces, in agreement with observations. In 
addition, the results showed that the chemistry is more sensitive to changes in the final 
density of the collapsing core in hot cores than in hot corinos. Finally, we find that ratios 
of deuterated sulphur bearing species in dense gas around hot cores and corinos 
may be good evolutionary indicators in a similar way as their non deuterated counterparts. 
\end{abstract}
\section{Introduction}
\label{intro}
Observations of local interstellar deuterated molecules have long been used to probe the physical 
conditions within interstellar clouds. Although the interstellar ratio of deuterium to 
hydrogen (D/H) atoms within the Milky Way is only $\sim 1.5 \times 10^{-5}$ \citep{lin95,oli03}, 
the observed degree of deuterium fraction\footnote{The deuterium fraction is the abundance 
ratio of a molecule containing a deuterium atom (XD) to the equivalent molecule with a hydrogen 
atom (XH). Deuterium fractionation is the process which leads to the enhancement 
of the deuterium fraction.} 
is enhanced in many astrophysical regions such as cold, dark cores (e.g. \citealt{tin20, crap05}), 
hot cores around massive star forming regions (e.g. \citealt{ehr20,font08}), lukewarm cores \citep{sak09}, 
and hot corinos around low mass star forming regions (e.g \citealt{cec98c,cec01,cec07}). 

Although there are only few detections of HD, the simplest deuterated species 
(e.g. \citealt{wri99, bert99, pol02, caux02, neu06, yuan012, ber013}), there is a growing body of 
observations, towards low- and high- mass star forming regions, for mono- as well as multiply 
deuterated species, such as H$_2$D$^{+}$  (e.g. \citealt{star99, star04, pil012}), DCN (e.g. \citealt{wils73, 
vand95}), HDO (\citealt{hen87, vand95}), and DCO$^{+}$ (\citealt{pen79, but88}). The first detection of a doubly deuterated 
species (D$_2$CO) was towards Orion KL by \citet{tur90}. The same molecule was then detected towards the low-mass protostar 
IRAS16293-2422 \citep{cec98c} with a D$_2$CO/H$_2$CO ratio 15 times higher than that obtained for Orion. 

To-date, around 32 deuterated species have been detected in a variety of interstellar clouds including ND$_3$, CHD$_2$OH, 
CD$_3$OH, D$_2$S, D$_2$O, HD$_2^{+}$, CH$_2$DOH, H$_2$D$^{+}$, C$_3$D, C$_4$HD, DC$_5$N, and DCOOCH$_3$ (e.g. see reviews 
\citealt{cec07, her09, cas012, tiel13} and references therein). 
Table (\ref{tab:list1}) lists the observed deuterated species in cores around massive (H) and low-mass (L) star forming 
regions only. 
\begin{table}
\centering
\caption{List of observed deuterated species in massive `H' and low-mass `L' star forming regions 
and the corresponding references.}
\label{tab:list1}
\begin{tabular}{lll} \hline
{\bf Species} & {\bf Region} & {\bf Ref (e.g.)}\\
\hline
{\bf H$_2$D$^{+}$}& H &  \citet{pil012}\\
& L& \citet{star99}\\
%
%
{\bf N$_2$D$^{+}$}& H& \citet{font06}\\ 
& L& \citet{emp09}\\ 
{\bf DCO$^{+}$}&H &\citet{pen79} \\ 
 &L &\citet{but88} \\ 
{\bf CH$_2$D$^{+}$}&H &\citet{rou13} \\ 
{\bf DCN} & H & \citet{wils73}\\
& L& \citet{vand95}\\ 
{\bf$^{\dag}$ DNC}& H & \citet{rod96}   \\ 
&L & \citet{vand95}   \\ 
{\bf HDS }&L &\citet{vand95}\\
{\bf D$_2$S }&L &\citet{vas03} \\
%
%
%
{\bf HDCO}&  H & \citet{lor85}\\
&  L& \citet{vand95}\\ 
{\bf D$_2$CO} &H &\citet{tur90} \\
& L  &\citet{cec98c} \\
{\bf CH$_3$OD}& H& \citet{sai94}\\
&L & \citet{par02}\\
{\bf CH$_2$DOH}&H & \citet{jacq93}\\
& L & \citet{par02}\\
{\bf CHD$_2$OH}&L  & \citet{par02}\\
{\bf CD$_3$OH}&L & \citet{par04}\\ 
{\bf C$_2$D }& H& \citet{vrt85}\\
&L & \citet{vand95}\\ 
{\bf C$_3$D }& L &\citet{sak09}\\ 
{\bf C$_4$D}& L  & \citet{sak09}\\
{\bf C$_3$HD} & L&\citet{sak09} \\ 
{\bf DC$_3$N}&L & \citet{sak09}\\
{\bf DC$_5$N}&L & \citet{sak09} \\
{\bf NH$_2$D }&H & \citet{rodk78}\\
&L & \citet{sha01}\\
%
%
{\bf ND$_3$}&L &\citet{vant02} \\ 
{\bf HDO}& H& \citet{hen87}\\
& L& \citet{vand95}\\ 
{\bf D$_2$O}&H& \citet{neil13a}\\
&L& \citet{but07}\\
{\bf CH$_2$DCN}& H & \citet{ger92}\\
{\bf DCOOCH$_3$}& L& \citet{dem10}\\
 \hline \hline
\end{tabular}
\flushleft
$\dag$ Data for massive star forming regions `H' is taken from Table 1 in \citet{rod96}
\end{table}

In parallel to observational studies, theoretical modelling of deuterium chemistry 
has also being taken place over the years. Early attempts at studying deuterium chemistry 
used simple gas-phase models \citep{wats80} or surface chemistry \citep{tiel83}. Both models 
were relatively successful in reproducing the observed deuterated species, at that time, 
and showed the role of grain surface reactions in enhancing deuteration of some species such 
as H$_2$CO. The first gas-grain chemical models were later developed by \citet*{bro89a,bro89b}. 
They found that surface chemistry can produce small but significant amounts of multi-deuterated molecules. 
Following this study, \citet{mil89} used a detailed numerical pseudo-time-dependent gas-phase chemical model 
to study deuteration in dense clouds. Their results showed that in cold clouds ($\sim$ 10 K) the main 
sources of fractionation are H$_2$D$^{+}$ and its daughter ions DCO$^{+}$ and H$_2$DO$^{+}$, while for warmer 
regions, up to 70 K, CH$_2$D$^{+}$, C$_2$HD$^{+}$, and associated species led to the high fractionation. 
The methodology of extending chemical networks to include deuterium was presented by \citet*{rod96} in 
their model of the deuterium chemistry in hot cores. 
They based their approach on the assumption that both H- and D-bearing species react with the same species 
with the same rate coefficient. Where there is an uncertainty, statistical branching ratios were assumed. 
Their model showed that the deuterium fractionation would be preserved in hot cores for at least 10$^{4-5}$ yrs 
after evaporation. Following  \citet*{rod96}, \citet*{rob20a, rob20b} developed new chemical models including, 
for the first time, the deuterated sulphur-bearing species and gas-phase chemistry of some doubly-deuterated 
species to investigate the influence of varying a wide range of physical parameters on the fractionation in 
interstellar clouds. They found that the fractionation ratios are temperature and fractionation process ( i.e. 
due to H$_2$D$^{+}$, CH$_2$D$^{+}$ or C$_2$HD$^{+}$) dependent. In addition, they showed that H$_2$S and HDS 
are good probes for regions where grain surface chemistry is important. 
They also commented that HDCO and D$_2$CO are good probes for fractionation on grain surfaces. 
The detection of multiply deuterated isotopologues of H$_3^+$ (e.g. \citealt{cas03, vas04}) motivated \citet{rob03, rob04} 
to release a new version of their model in which multiply deuterated isotopologues of H$^{+}_{3}$ are taken into account. 
Their results revealed that the inclusion of D$_2$H$^{+}$ and D$^{+}_{3}$ enhances the fractionation of ionic and 
neutral species because it allows more deuterium to transfer to other species in dark clouds.
In addition, experimental work and chemical models showed that, in cold environments, the spin state of 
species (ortho, para, meta) affects the deuterium fractionation (e.g. \citealt{gerl02, wal04, flo06, hugo09, sip10}), 
by allowing ortho-H$_{2}$ to react with deuterated isotopologues of H$^{+}_{3}$ and reducing deuterium fractionation. 

It is clear that deuterium fractionation is a function of the chemical as well as the physical conditions of star 
forming regions. This paper is dedicated to investigate, theoretically, whether the evolution of deuterium chemistry 
differs significantly between low- and high- mass star forming regions by modelling deuterium chemistry under physical 
conditions likely similar to those of hot cores and corinos. 
Our model treats the desorption of the species, deuterated and non-deuterated, to be temperature dependent during the protostellar 
phase (the warming up phase). This approach allows for better identification of chemical tracers for various evolutionary stages in regions 
where the protostar has started to affect the surrounding gas and dust.  

The present model, UCL\_DCHEM, is an adaptation of the UCL\_CHEM model \citep{viti99, viti04}, which in its original form did not include 
a deuterium network.\footnote{This paper is an extension with several updates to the original, unpublished, study performed by Zainab Awad as part of her Ph.D; see \citet{awadphd}.}

This paper is organized as follows: in \S \ref{model} we describe our model; 
in \S \ref{res} we present the results of the grid of models we ran. We compare 
our model calculations with observations and other models in \S \ref{comp}, 
and give our conclusions in \S \ref{conc}.
\section{The Model}
\label{model}
We have used a time-dependent gas-grain chemical model described in details in 
\citet{viti04} and \citet{awad10}. The model is a two-phase code. The first phase (Phase I) 
simulates the free-fall collapse of a core as described in \citet {raw92}. It starts 
with diffuse, mainly atomic material which has an initial number density of $\sim 400$ cm$^{-3}$ 
and temperature of 10 K. The material undergoes a free-fall collapse up to a given final density 
considered here as a free parameter. 
During phase I, gas-phase chemistry and freeze-out on to grain surfaces occur. Accreted atoms and molecules 
hydrogenate or deuterate when possible. The depletion (or freeze-out) efficiency is determined by the amount 
of the gas-phase material frozen onto the grains. 
Since CO is the most abundant species, after H$_{2}$, and because its depletion percentage is an important 
factor in measuring the deuterium fractionation (e.g. \citealt{cas02b, bacm03}), we based our calculation of the 
depletion efficiency on the abundance of CO species. The depletion efficiency is regulated by varying the sticking 
probability of gas-phase species onto grain surfaces \citep {raw92}. In Phase II, we follow the chemistry 
of the remnant core, after the star is born. In this phase, the central star heats up the surrounding gas and dust, 
causing selective sublimation of the icy mantles. We adopt an identical treatment for the time dependent ice 
sublimation to \citet{viti04}.

In this work, we have extended the species set used in \citet{awad10} by including all the possible mono-
deuterated counterparts for H-bearing species to model deuterium chemistry. The only doubly deuterated species 
included in this model is D$_2$CO and its parent ion HD$_2$CO$^{+}$. 
The exclusion of other doubly-deuterated species, as well as triply-deuterated species is of course a limitation 
of this model. However we note that the aim of this work is to characterize the general trends of the chemical evolution 
of hot cores and corinos (hence gas at temperature higher than 100K). In these warm  regions deuteration is not as efficient 
as in dark clouds. The deuteration process is driven by H$_{2}$D$^+$, C$_2$HD$^+$ and/or CH$_2$D$^+$ ions that are formed via 
radiative association reactions involving H$_3^+$, C$_2$H$_2^+$, and CH$_3^+$ ions, respectively. At low temperatures, H$_3^+$ 
is the most abundant ion and hence fractionation of H$_2$D$^+$ is the most important of the three. At temperatures higher 
than 25K, however, H$_2$D$^+$ is destroyed by H$_2$. CH$_2$D$^+$ and C$_2$HD$^+$ are important for deuteration in regions 
colder than 60 and 80 K, respectively (\citealt{mil20, par09}). 
Hence, the amount of doubly and triply deuterated forms of H$_3^+$ or CH$_3^+$ should be negligible at 
temperatures higher or equal to 100K. While therefore our chemistry is limited during Phase I (leading to the warm up phase 
starting with a deuteration fraction that may not be as accurate) by focusing on the trends during the warm up phase, in the 
warm gas, a slightly lower or higher deuteration at the beginning of Phase II should not affect the trends. 
In fact chemical models by \citet{rob03} 
revealed that the inclusion of multiply deuterated forms of H$_3^+$ improved the results for certain ions, 
namely N$_2$H$^+$. Our model supports this result (see Section 3).

Our chemical network is based on the network previously described by \citet{rob20a}. However, we 
updated several reactions following \citet{woo07}\footnote{This work was performed 
before the release of the UMIST 2012 ratefile \citep{mce013}.} ratefile and using \citet*{rod96} recipe in 
generating the reactions involving the deuterium counterparts. Moreover, the rate coefficients for some radiative 
association reactions and the binding energies for surface species were also updated following \citet{rob04} 
and \citet{rob07}, respectively. 
Beside these updates, we included all the freeze-out reactions for hydrogen 
bearing species and their deuterium counterparts, assuming that the products 
will have the same branching ratios adopted for their hydrogen equivalents.
The surface chemistry considered here is simple in that it
includes, besides the H$_2$ and HD formation on grains, rapid hydrogenation of 
species, where energetically possible. Apart from direct hydrogenation, the only other surface 
reactions we include are the formation of methanol from CO and of CH$_3$CN from the reaction of 
methane, CH$_4$, with HCN, as it has been shown that gas phase reactions are not sufficient to 
form these two species (e.g. \citealt{tiel82,wat03,gar08}). As a first approximation and according 
to experimental results of the thermal desorption non-deuterated species from icy mantles (e.g. \citealt{col03a, 
col04}), we assumed that deuterated species will desorbe at the same temperature recorded for their hydrogen 
counterparts ({\it McCoustra - private communication}). 
Desorption temperatures are listed in Table (2) in \citet{viti04} for hot cores and determined by \citet{awad10} for hot corinos.

This work studies the influence of changing the final density of the collapsing core, the depletion 
efficiency of the gaseous species onto grain surfaces, and the effect of varying the 
collapsing mode (free fall or retarded)\footnote{A retarded collapse means that the speed of 
the collapse is a factor of that of the free-fall collapsing speed which we assume it is unity. 
In this model we change the speed of the collapse by varying the collapse parameter defined in the 
modified collapse equation by \citet{raw92}.} on the chemical evolution of deuterated species around 
low-mass (1 M$_{\odot}$) and higher mass (5 \& 25 M$_{\odot}$) protostars. 

In the first instance, we ran a total of 9 models, where the initial elemental abundances and parameters 
used for the grid are listed in Tables (\ref{tab:initial}) and (\ref{tab:grid2}). 
Our chemical network consists of 265 species (including 92 deuterated species and 60 surface species) 
linked in 4204 reactions both in gas-phase and on grain surfaces. 

\begin{table*}
   \centering
\caption{Initial elemental abundances, with respect to the total number of hydrogen nuclei, 
and physical conditions assumed in our model (taken from \citealt{viti04, awad10}). Note that 
for the described parameters, hereafter, the temperature is the gas kinetic temperature (in K) and 
the density is the gas number density (in cm$^{-3}$).}
  \label{tab:initial}
    \leavevmode
    \begin{tabular}{lll} \hline
\multicolumn{3}{c}{Initial elemental abundances} \\ \hline 
Carbon & \multicolumn{2}{c}{$1.79 \times 10^{-4}$} \\
Oxygen & \multicolumn{2}{c}{$4.45 \times 10^{-4}$} \\
Nitrogen & \multicolumn{2}{c}{$8.52 \times 10^{-5}$} \\
Sulphur & \multicolumn{2}{c}{$1.43 \times 10^{-6}$} \\
Helium & \multicolumn{2}{c}{$7.50 \times 10^{-2}$} \\
Magnesium & \multicolumn{2}{c}{$5.12 \times 10^{-6}$} \\ \hline 
\multicolumn{3}{c}{Physical parameters} \\ \hline 
{\bf The parameter} & {\bf Hot Corino} &  {\bf Hot Core}\\
& \citet{awad10}&\citet{viti04}\\ \hline 
$^{\dag}$Core density (cm$^{-3}$) & $10^{7} - 10^{8}$ & $10^{6} - 10^{7}$ \\
Core temperature (K) & 100 & 300 \\
Core radius & 150 AU & 0.03 pc\\
Protostellar Mass (M$_{\odot}$)& 1 & 5 $\&$ 25\\ 
&\multicolumn{2}{c}{\underline {\bf For Both Models}} \\
$^{\dag}$Depletion percentage (\%)& \multicolumn{2}{c}{85 - 100} \\ 
$^{\dag}$The collapsing mode & \multicolumn{2}{c}{free fall - retarded} \\ \hline \hline
    \end{tabular}
\flushleft
$^{\dag}$ This parameter varies only during the collapsing phase (Phase I).\\
\end{table*}
\begin{table*}
\centering
\caption{Summary of the grid of our models in this study as described in \S\ref{res}} %
\label{tab:grid2}
\begin{tabular}{lllll} \hline
{\bf Models } & {\bf Mass}& {\bf Temperature}&{\bf $^{\dag}$Density}&{\bf $^{\dag}$Depletion}\\
              & {\bf M$_{\odot}$ }&{\bf K }&{\bf cm$^{-3}$ }&{\bf $\%$}\\ \hline 
{\bf M1} & 1 & 100 &  2.0$\times$10$^{8}$ & 100 \\
{\bf M2} & 1 & 100 &  1.0$\times$10$^{7}$ & 100 \\
{\bf M3} & 1 & 100 &  2.0$\times$10$^{8}$ & 85 \\ \hline
{\bf M4} & 5 & 300 &  1.0$\times$10$^{7}$ & 100 \\
{\bf M5} & 5 & 300 &  1.0$\times$10$^{6}$ & 100\\
{\bf M6} & 5 & 300 &  1.0$\times$10$^{7}$ & 85 \\ \hline
{\bf M7} & 25 & 300 &  1.0$\times$10$^{7}$ & 100 \\
{\bf M8} & 25 & 300 &  1.0$\times$10$^{6}$ & 100 \\
{\bf M9} & 25 & 300 &  1.0$\times$10$^{7}$ & 85 \\ \hline 
&\multicolumn{3}{c}{$^{\star}$Models with different collapse speeds} &\\ \hline
{\bf Models } & {\bf speed of collapse} & \multicolumn{3}{c}{\bf Notes} \\ \hline
{\bf ff}      & 1              & \multicolumn{3}{l}{free-fall collapse with assumed speed unity.} \\ 
{\bf ret 0.5} & 0.5 ff & \multicolumn{3}{l}{retarded collapse with half the speed of the ff model.} \\ 
{\bf ret 0.1} & 0.1 ff & \multicolumn{3}{l}{retarded collapse with tenth the speed of the ff model.} \\ \hline 
\end{tabular} 
\flushleft
M1, M4 and M7 are the reference models in the grid.\\
$^{\dag}$ These parameters vary only during the collapsing phase (Phase I) of the chemical model.\\
$^{\star}$ These models have similar physical conditions to that of model M1. ff: free-fall collapse model, ret: retarded collapse model
\end{table*}
\section{Results and Discussion}
\label{res}
\begin{figure*}
\begin{center} 
\includegraphics[trim=1cm 0.5cm 2cm 1.5cm, clip=true, width=18cm]{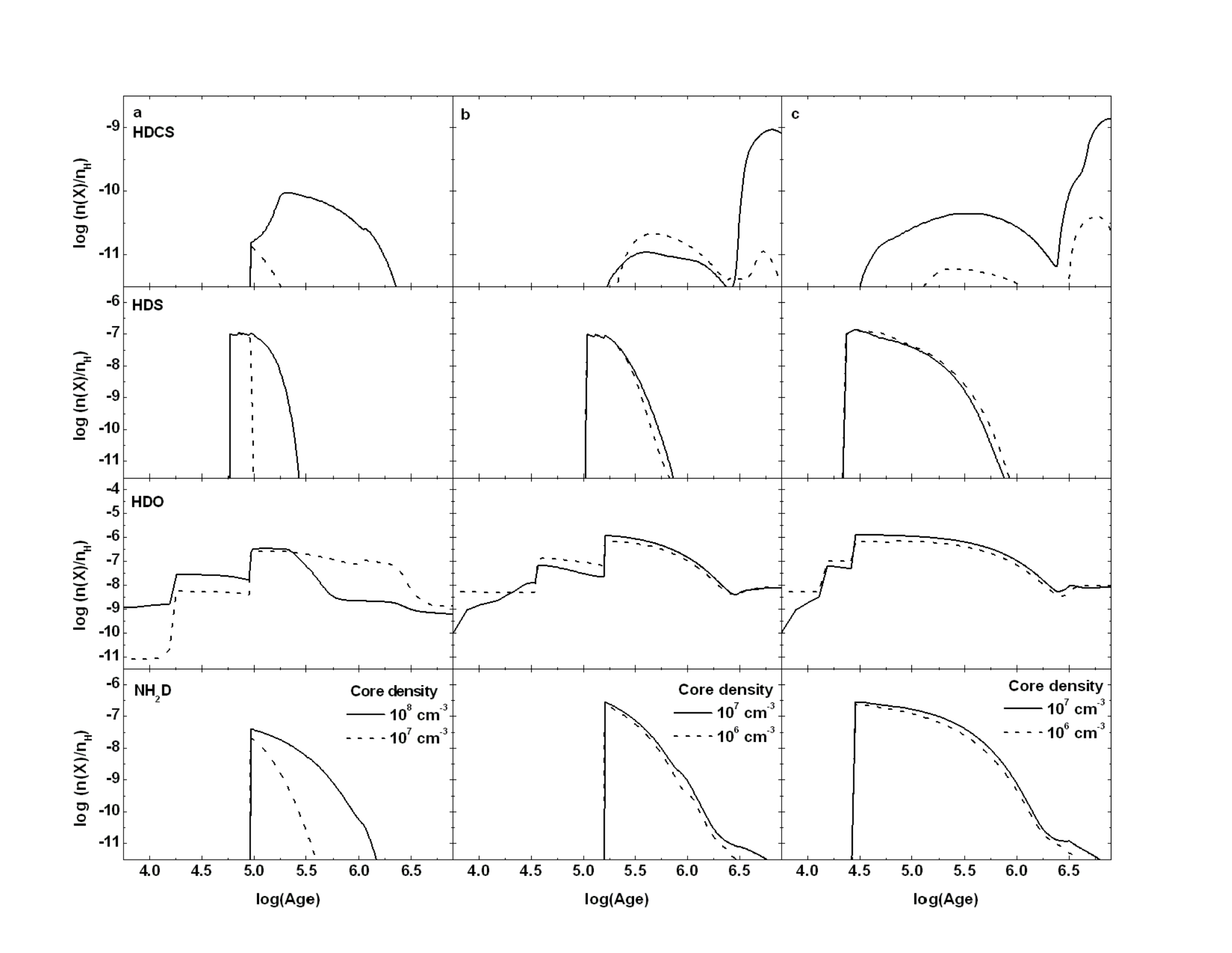} 
\caption{The chemical evolution (from top to bottom) of HDCS, HDS, HDO, and NH$_2$D in hot corinos 
(panel a: 1 M$_{\odot}$) and hot cores (panel b: 5 M$_{\odot}$, and panel c: 25 M$_{\odot}$) as a 
function of time, for Phase II calculations. The different curves compare the evolution of the species at two 
different final densities for the collapsing cloud (see key in bottom plots, and Table \ref{tab:grid2}).}
\label{fig:1}
\end{center}
\end{figure*}

\begin{figure*}
\begin{center} 
\includegraphics[trim=1cm 0.5cm 2cm 1cm, clip=true,width=18cm]{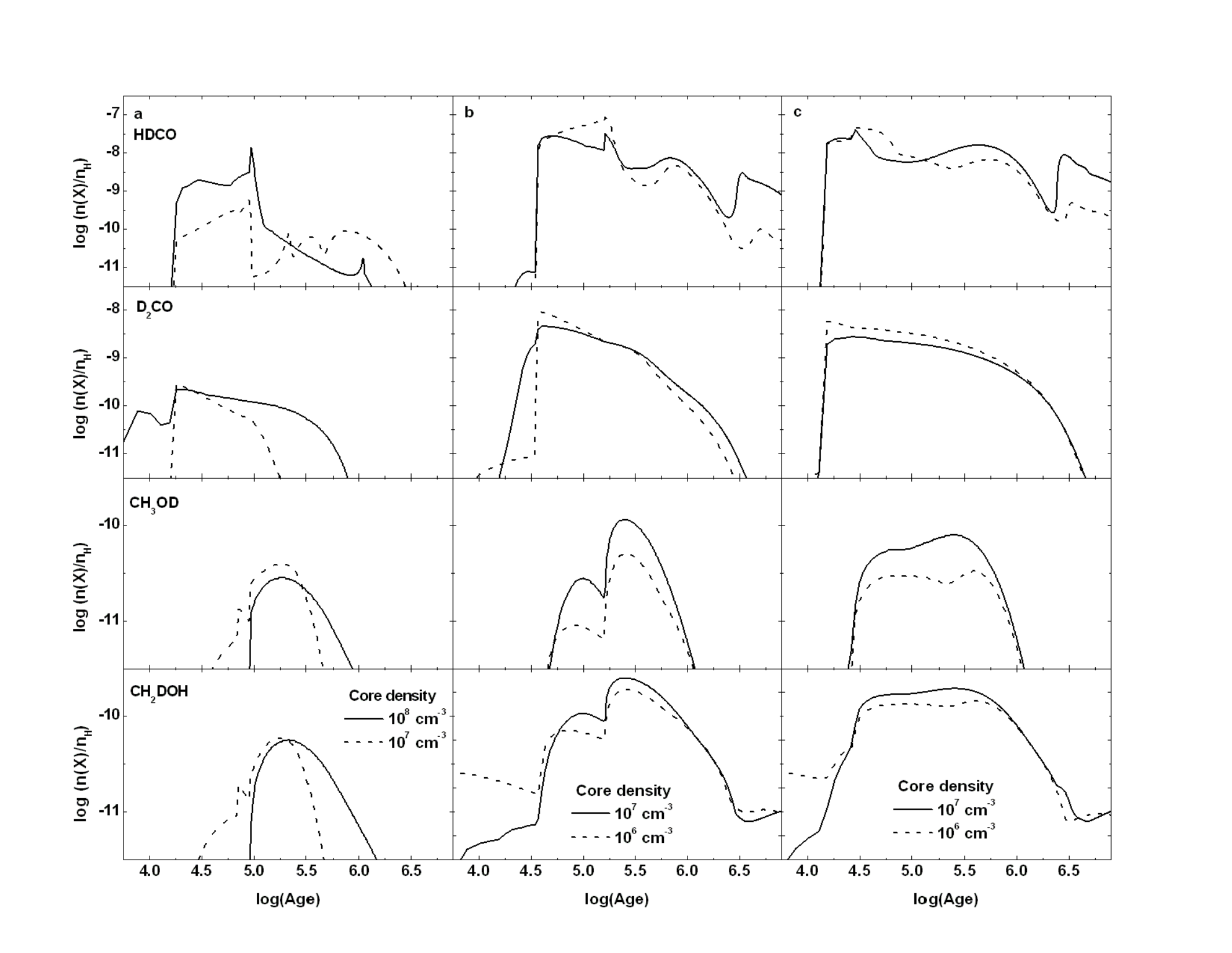} 
\caption{Similar to Fig. \ref{fig:1} but for deuterated formaldehyde and methanol (see key).}
\label{fig:2}
\end{center}
\end{figure*}

We investigate the sensitivity of deuterium chemistry to changes in the physical conditions in low- and high-mass star 
forming regions by comparing the reference models M1, M4 and M7 (for masses: 1, 5, and 25 M$_{\odot}$; respectively 
- see Table (\ref{tab:grid2})) to the following models:  M2, M5, and M8 to study the influence of changing the final 
density of the collapsing cloud; and models: M3, M6, and M9 to explore the influence of varying the depletion percentage 
of gaseous species onto grain surfaces. Note that the mass in the second column is that of the newly formed star. 
The influence of changing the final density of the collapsing core and the depletion efficiency were 
studied by adopting some standard values - see Table (\ref{tab:grid2}) - and then decreasing their values by 10 
and 15\%, respectively. 
The standard values are listed in Table (\ref{tab:grid2}): model M1 for a solar mass hot corino and models M4 and 
M7 for 5 and 25 solar masses hot cores. The choice of these values is arbitrary providing that the model physical conditions 
are within the observed ranges for low- and high- mass cores. Figs. (\ref{fig:1} - \ref{fig:4}) show the predicted 
fractional abundances as a function of time for hot corinos (panel a) and hot cores (panels b and c). 

The effect that the collapse timescale may have on the chemistry has been previously discussed for low-mass stellar 
cores by \citet{sak08}. Hence here we focus on the influence of varying this collapse timescale for hot corinos only:
We compare models `ret 0.5' and `ret 0.1' to the standard `ff' model of a hot corino to understand the impact of 
varying the speed of the collapse on the chemical timescales of the region under study. 
Fig. (\ref{fig:5}) illustrates the fractional abundances of species in a hot corino as a function of time for 
free-fall (solid line) and retarded collapse (dashed and dotted line) models.
\subsection{Sensitivity to the final core density}
\label{df}
Generally, a change in densities affects hot corinos more than hot cores. This result is shown in panels (a), (b) 
and (c) in Figs. (\ref{fig:1}) and (\ref{fig:2}). These figures represent Models M1(M2), M4(M5), and M7(M8), 
when the final density is 10$^8$(10$^7$) cm$^{-3}$ for hot corinos, and 10$^7$(10$^6$) cm$^{-3}$ for hot cores.

As expected, species in hot corinos are more abundant in denser cores (solid line) than in less dense 
cores (dashed line); exceptions are CH$_3$OD, CH$_2$DOH, and HDO that show higher abundances for less dense regions, in particular 
during early times (t $\le$ 10$^{5}$ yrs). Generally, the two deuterated methanol counterparts show 
the same evolutionary profile, but CH$_3$OD seems to survive longer than CH$_2$DOH. 
The chemical analysis of those species, in dense 
regions (10$^8$ cm$^{-3}$) at early times, reveals that they are  
destroyed via reactions with H$^+$ that are more efficient at higher densities. Therefore, their abundances 
remain high for longer times in less dense regions. For both models, M1 and M2, HDO is formed via reactions 
involving H$_2$CO (see next section) which are more efficient at lower densities (Model M2). 
The abundance of HDO is enhanced for times $\geq$ 10$^{5}$ yrs, which is consistent with the desorption times of 
H$_2$CO from grain surfaces in the H$_2$O co-desorption event \citep{awad10}. Moreover, the number of destruction 
pathways of heavy water in denser cores is larger than that for less dense cores, allowing more HDO to remain in the 
gas-phase.

All the species experience a rapid and steeper decline in their abundances in less dense (10$^7$ cm$^{-3}$) 
hot corinos (Panel a) than in denser ones (10$^8$ cm$^{-3}$). 
The chemical trends in the two models remain the same except for HDCO, which shows fluctuations in its 
abundance for low density hot corinos, in particular after 10$^{5}$ yrs. 
The chemical analysis of HDCO at this particular time for a lower density hot corino reveals that the 
species is involved in a larger number of gas-phase reactions than in the case of a denser core. 

Molecules in massive cores (with higher temperature) are less affected by the decrease in core density. 
However, deuterated water shows a slightly higher abundance at times 
$\le$ 2 $\times$ 10$^{5}$ yrs, and both forms of deuterated methanol are less abundant in lower 
density cores. 
\subsection{Impact of varying the depletion efficiency}
\label{fr}
Observations reveal that the enhancement of the deuterated species to their fully 
hydrogenated forms arises in regions where the CO molecules are heavily depleted onto dust grains 
(e.g. \citealt{cas02b, bacm03, mil05, crap05, chen011}). Therefore, we studied the influence of varying the 
depletion percentage on the fractional abundances of the species in hot corinos and cores. 

\begin{figure*}
\begin{center} 
\includegraphics[trim=1cm 0.5cm 1cm 1cm, clip=true,width=18cm]{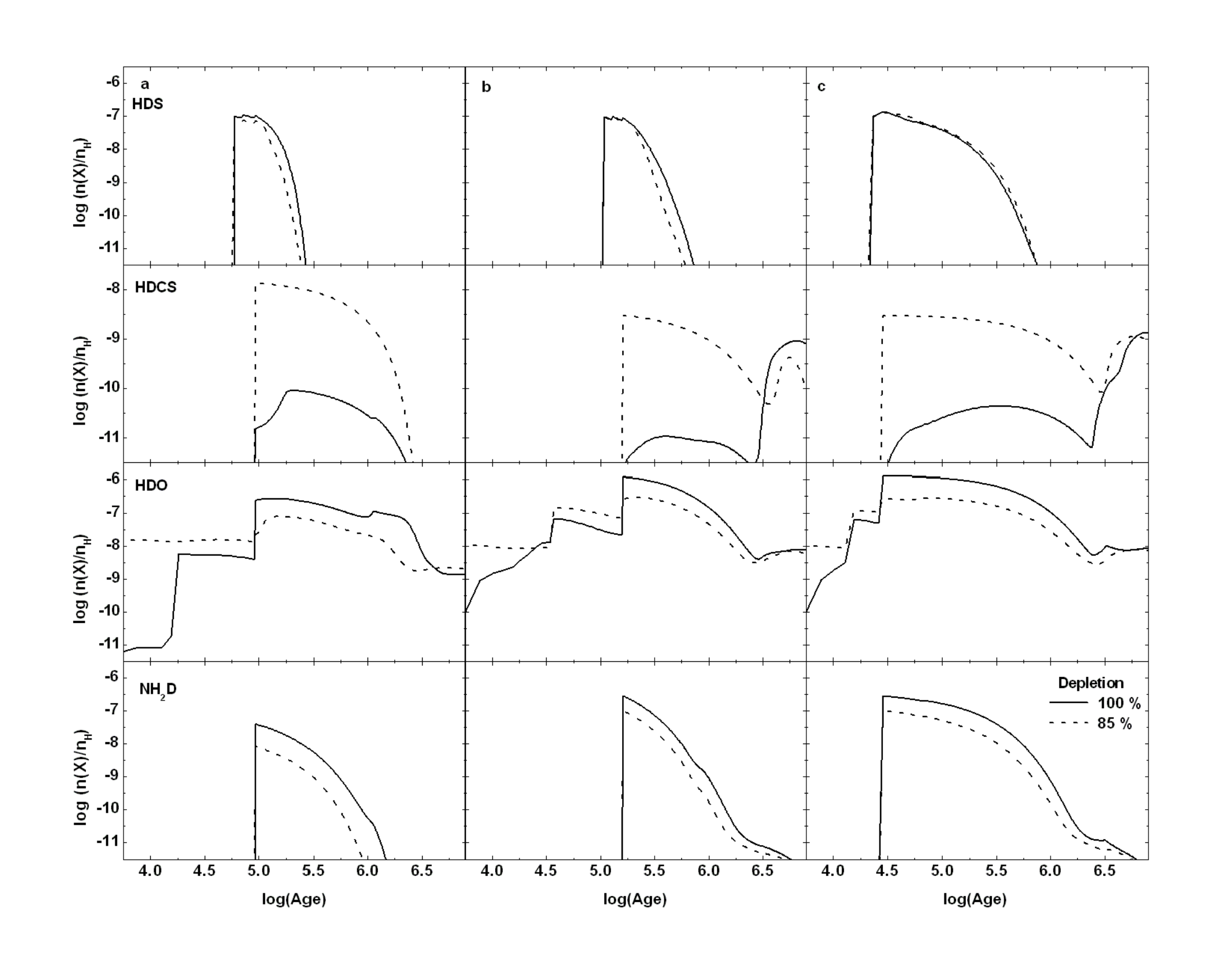} 
\caption{Chemical evolution of HDCS, HDS, HDO, and NH$_2$D as a function of time, during Phase II, at 
different depletion on grain surfaces (see key text in bottom right plot, and Table \ref{tab:grid2}) 
for different cores: (a) 1 M$_{\odot}$, (b) 5 M$_{\odot}$, and (c) 25 M$_{\odot}$.}
\label{fig:3}
\end{center}
\end{figure*}

\begin{figure*}
\begin{center} 
\includegraphics[trim=1cm 0.5cm 1cm 1cm, clip=true,width=18cm]{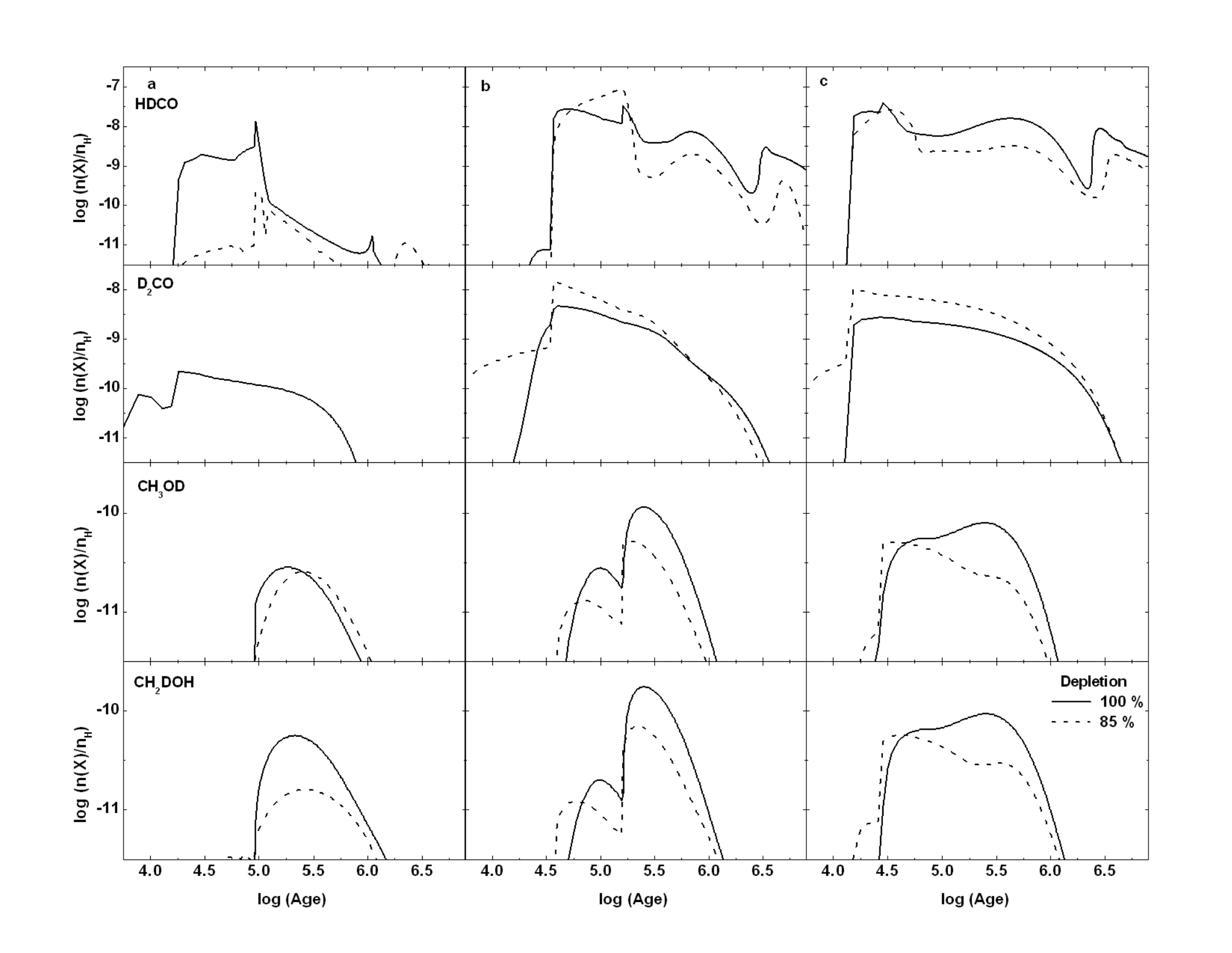} 
\caption{Similar to Fig. \ref{fig:3} but for deuterated organic species (see key).}
\label{fig:4}
\end{center}
\end{figure*}

\begin{figure*}
\begin{center} 
\includegraphics[trim=0cm 1cm 0.5cm 1.5cm, clip=true,width=15.5cm]{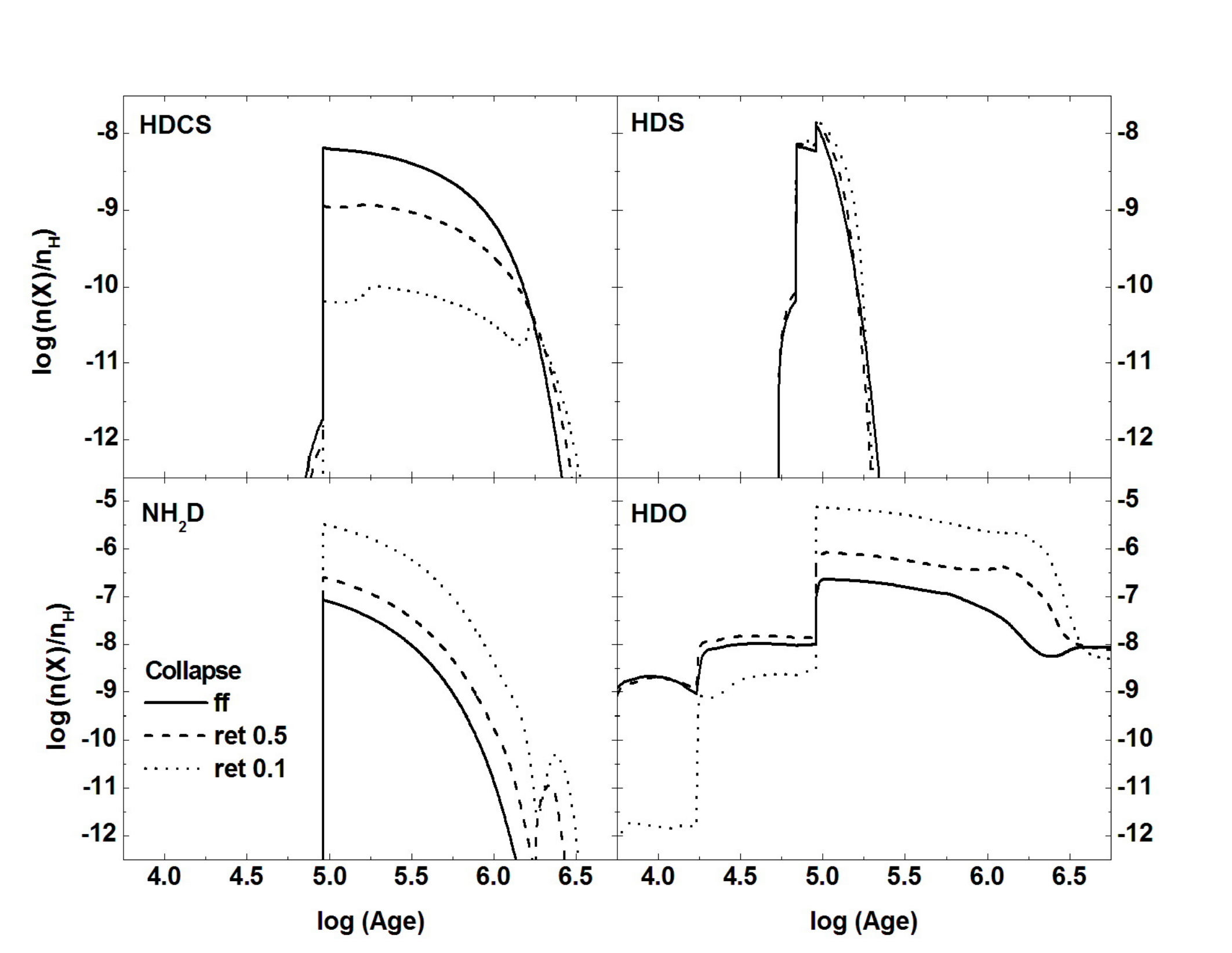} 
\includegraphics[trim=0cm 0cm 1cm 1.5cm, clip=true,width=15cm]{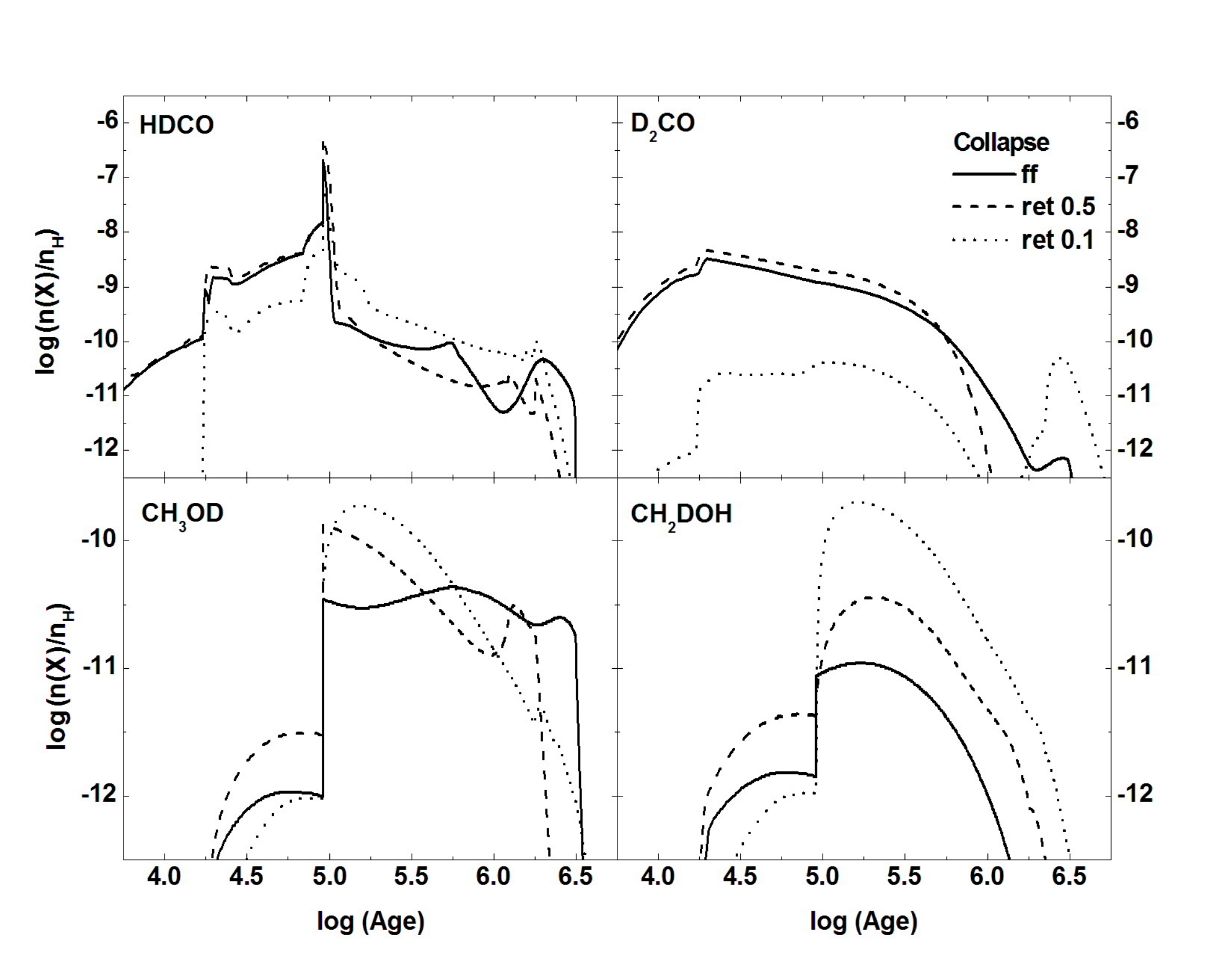} 
\caption{Chemical evolution of a selected set of deuterated species during the 
the warming-up phase (Phase II) as a function of time using different collapsing modes: 
the free fall (ff: solid line), retarded; 0.5ff speed (ret 0.5: dashed line) 
and 0.1ff speed (ret 0.1: dotted line), see key and Table \ref{tab:grid2}.}
\label{fig:5}
\end{center}
\end{figure*}

In this section, we discuss the results of modelling hot corinos and cores with fully (Models M1, M4, \& M7)
and partially depleted gas (Models M3, M6, \& M9), respectively.
The chemical evolution of deuterated molecules in a fully depleted gas 
(solid line) is plotted in comparison with the case of partially depleted gas (dashed line), 
both as a function of time, in Figs (\ref{fig:3}) and (\ref{fig:4}). 
The term `fully depleted' refers to interstellar gas which is close to 100\% freeze-out by the end of the 
collapsing phase (Phase I), while `partially depleted', in this paper, means an 
arbitrary freeze out percentage of $\sim$ 85\%.
Generally, models with partially depleted gas yield lower abundances for all deuterated species, apart from HDCS 
which is enhanced between 9$\times$10$^{4}$ yrs and $\sim$ 1 million years. In massive cores, and for times 
earlier than 1.6$\times$10$^{5}$ yrs, the fractional abundances of HDO, D$_2$CO, and the two deuterated 
forms of CH$_3$OH show a slight enhancement with lower depletion percentage. 
The abundance of HDS is the least affected by a lower depletion percentage. 
This is probably due to the fact that HS and its deuterated counterpart are not formed or enhanced on the icy mantles. 
Chemical analysis of HDCS reveals that prior to 9$\times$10$^{4}$ yrs, the formation 
rate of HDCS is similar in both models while after that time, and when mantle 
species sublimate, the formation rate of HDCS increases in models with partial depletion. 
This result may indicate that this molecule is mainly formed via gas-phase reactions and is 
not a mantle species. 
HDO, in hot corinos (Fig. \ref{fig:3}, panel a), with higher depletion percentage (solid line) 
shows higher abundances at late times. At times of $\sim$ 1.7 $\times$ 10$^{4}$ yrs, the abundance of 
HDO shows a sudden increase which cannot be explained in terms of mantle sublimation. 
The chemical analysis at the time around this `jump' reveals that 
it is caused by the presence of methane (CH$_4$) in the gas phase. 
The time of the `jump', $\sim$ 1.7 $\times$ 10$^{4}$ yrs, is associated with that observed previously for formaldehyde 
\citep{awad10}. 
In this model, HDO is produced via reactions involving H$_{2}$CO or HDCO with H$_2$DO$^+$. As a consequence, we argue 
that the increase in the HDO abundance can be explained as a result of the enhancement in the formation of either 
H$_{2}$CO or HDCO, as follows. CH$_4$ evaporates from grain surfaces at $\sim$ 1.7 $\times$ 10$^{4}$ yrs, then it 
undergoes many gas-phase reactions some of which form H$_2$DO$^+$ and CH$_3$. These two molecules are, in turn, used 
to form H$_2$CO and HDCO leading to an enhancement in their abundances. 
Therefore, the jump of HDO can be attributed, indirectly, to the evaporation of 
methane from grains surfaces.

On the other hand, D$_2$CO is enhanced in hot corino models with full depletion. This result is 
supported by \citet{bacm03}, see Section (\ref{comp}). 
In hot cores, the deuterated forms of H$_2$CO and CH$_3$OH (Fig. \ref{fig:4}) for both models 
are abundant for longer times than in hot corinos and hence these species may be good tracers 
of hot cores. For earlier times (i.e. t $< 10^4$ yrs), deuterated H$_2$CO and CH$_3$OH (mainly formed on grains) 
can be formed, in the gas phase, via reactions involving HCN (e.g. H$_2$DCO$^+$ + HCN $\rightarrow$ HCNH$^+$ + HDCO) 
and NH$_3$ (e.g. CH$_3$OHD$^+$ + NH$_{3} \rightarrow$ NH$_{4}^{+}$ +
CH$_3$OD). 

Unlike the case of hot corinos, D$_2$CO is observable in hot cores even when the 
depletion percentage is low. This molecule is efficiently formed via the reaction 
`HD$_2$CO$^+$ + H$_2$O', and is destroyed by H$_3$O$^+$ and HCO$^+$. The latter dominates 
the destruction routes at times later than 1.4 $\times 10^{4}$ yrs, but it is less efficient 
in hot core models with partial freeze out, leading to a higher D$_2$CO abundance.
HDCO is abundant during most of the hot core lifetime and therefore could be used as a good tracer 
for hot cores. It is interesting to note that HDCO shows a sudden increase in its abundance, even at low depletion efficiencies, around 
4 $\times 10^{4}$ yrs, which cannot be explained by its sublimation from grain mantles (as in the hot corinos case). 
This `jump' in abundance is related to the formation of CH$_4$, as explained above for hot corinos, but 
in hot cores H$_2$CO is mainly formed via oxidization reactions of CH$_3$, which is 
formed via the efficient destruction of CH$_4$. 

DCN is another species that seems to be affected by the degree of depletion in Phase I; 
\citet{par09} observed and
modelled the excitation of HCN and DCN in Clump 1 and 3 of the Orion Bar and estimated the DCN/HCN  ratios, on 
average, to be 0.7 and 1.1 (using the rotational diagram method) or 0.3 and 0.8 (using the LVG analysis), respectively. 
Our model calculations are in the range determined by \citet{par09} and indicates that, 
in general, fully depleted cores have a lower deuterium fraction 
($\sim 0.1 - 0.7\%$) than partially depleted cores ($\sim 0.8 - 1.2\%$).  
A chemical analysis shows that the rate of formation of DCN is higher in regions with full depletion 
compared to those with partial depletion. The reason for this is that for less depleted cores, DCN formation is 
mainly via the reaction `NH$_3$ + DCNH$^+$' which is inefficient in fully depleted cores. Moreover, 
DCN is extensively destroyed by the ion `N$_2$H$^+$' in cores with fully depleted gas. This  route is of equal 
weight with other destruction pathways in cores with partially depleted gas. 
Our results for fully depleted models are consistent with those obtained for Clump 1 while partial depletion results 
are in better agreement with the observed values in Clump 3. Our modelling suggest that 
the percentage of the DCN/HCN molecular ratio can be used to trace the initial level of depletion (i.e. before the 
protostar switches on) in a given region.

\subsection{Sensitivity to the mode of collapse}
\label{bc}
In their study of the chemistry of the low-mass core IRAS 04368+2557 in L1527, \citet{sak08} concluded that 
the timescale of the collapse of the pre-stellar core affects the abundances of the species in the region under study. 
The shorter the timescale (i.e. the faster the collapse) the more abundant the species are, in particular C-bearing 
species (C$_n$H$_m$). This result motivated us to run two extra models for a solar mass core in which we vary the speed 
of the collapse in Phase I. 
The results are then compared to those of a standard solar mass hot corino (Model M1 in this work which undergoes a free-fall 
collapse), see the parameters of Model M1 and other models in Table (\ref{tab:grid2}). 
Models with longer collapse timescale (e.g. Model ret 0.1) are expected to form more species than models 
with shorter collapse timescale (e.g. Model M1), and therefore show an enhancement in the fractional abundances 
of their molecular content in Phase II. 

Figure (\ref{fig:5}) shows the chemical evolution of the deuterated species during Phase II, for the free fall 
(solid line) and the retarded collapse models in which the speed of the collapse was decreased to half (dashed 
line) and a tenth of its free fall value (dotted line). 
From this figure, we note that when species are evaporated from the grain surfaces (after $\sim$ 4.8$\times$10$^{4}$ yrs): 
(1) the least affected species by changing the mode of collapse are HDS and HDCO, and 
(2) the fractional abundance of HDCS and D$_2$CO decrease while that of the rest of the species increases as the speed of the 
collapse decreases. The most affected species, and hence possible good indicators of the collapsing mode, are 
HDCS, NH$_{2}$, CH$_2$DOH, and HDO. We conclude that HDO/H$_2$O is affected by the collapse history of the pre-stellar core. 

Our results are in agreement with the findings of \citep{sak08} in which the affected species in models with longer collapse 
timescale (Model M1 - solid line) possesses the lowest abundance of all models. We hence conclude that the speed of the collapse 
influences the fractional abundances of hot corinos.
\subsection{Evolutionary stages indicators}
\label{indic}
\begin{figure*}
\begin{center} 
\includegraphics[trim=1cm 0cm 1cm 1.5cm, clip=true,width=18cm]{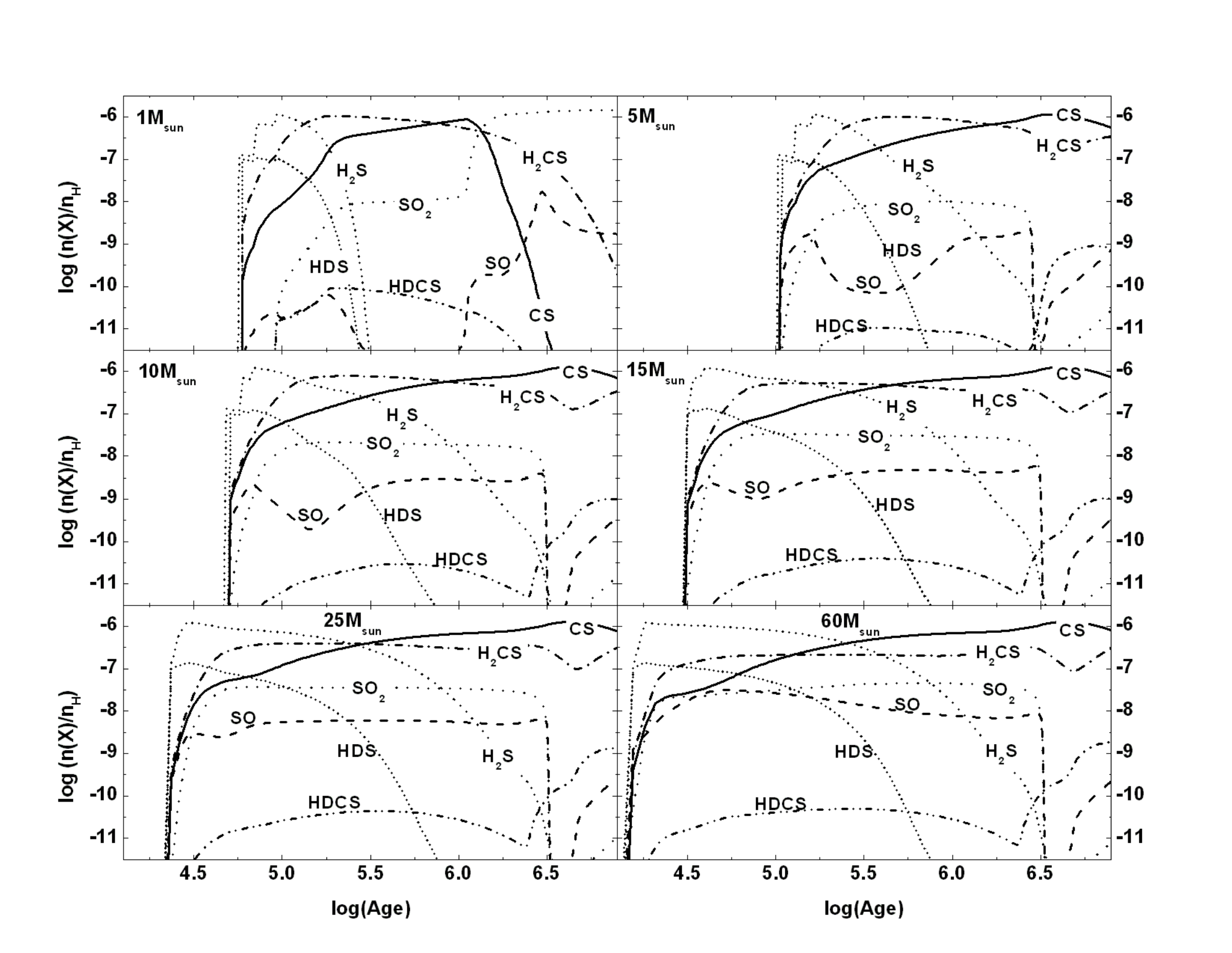} 
\caption{The chemical evolution of S-bearing species and their deuterated counterparts 
as a function of time in warm and hot cores for various masses as indicated on the plots 
(upper left corners). The fractional abundances of the studied species are represented 
by different line styles. } 
\label{fig:6}
\end{center}
\end{figure*}

\citet{viti04} studied the chemical evolution of hot cores around stars 
with various masses from 5 to 60 solar masses. They found that ratios of sulphur bearing 
species (e.g. H$_2$S/SO$_2$, H$_2$S/CS, SO/CS) are good indicators of the early stages 
of massive star formation while large organic molecules such as CH$_3$OH, 
HCOOCH$_3$, and C$_2$H$_5$OH indicate late evolutionary stages. In this work, 
we aim to investigate whether the deuterated counterparts of these evolutionary 
indicators can also be used for the same purpose. We, therefore,
ran a grid of four additional models to cover the range of stellar masses 
between 10 and 60 solar masses (as in \citealt{viti04}). 
Fig. (\ref{fig:6}) illustrates the chemical evolution of sulphur bearing species and their 
deuterium counterparts. Inspection of this figure shows that, as their non-deuterated counterparts,
the ratios of deuterated sulphur bearing species may be good chemical evolutionary tracers of hot 
cores in all cases explored. 
\section{Comparison with observations and other model calculations}
\label{comp}
We briefly discuss our results by qualitatively comparing them to one representative hot corino and one hot core in \S \ref{comp-obs}, 
and by comparing our models with previous chemical modelling in \S \ref{comp-mod}.
\subsection{Comparison with Observations}
\label{comp-obs}
Our physical conditions for 
models of hot corinos (model M1) are comparable to those observed for the IRAS 16293-2422 source, while those of hot cores (model M7) 
are close to observations of Orion KL hot core. Hence we briefly compare our results with the observed abundances of these two sources.\\
{\it \bf  IRAS 16293-2422} is a nearby Class 0 protostar at distance $\sim$ 120 pc and predicted age 
of 10$^5$ years \citep{and93}. It has an inner small (150 AU) condensed ($\sim 2\times10^{8}$ cm$^{-3}$) region known 
as `hot corino' \citep{cec98c}. 
A comparison between Model M1 fractional abundances and the calculated molecular D/H ratios of various species 
with those derived from observations of IRAS 16293-2422 source and the times of 
best fit of our calculations with observations are summarized in Table (\ref{tab:obs}). 
All of our fractional abundances are given relative to the total number of H nucleons in the region. 
Note that for most observations quoted in Table (\ref{tab:obs}) we cannot distinguish the emission 
from the inner hot corino region and that from the external outer region of the source; an exception is 
the deuterated water observations by \citet{cout13}. 

\begin{table*}
\centering
\caption{Comparison between observations of deuterated species and molecular D/H ratios 
in the IRAS 16293-2422 hot corino source and our model M1 calculations of Phase II.}
  \label{tab:obs}
    \leavevmode
    \begin{tabular}{lllll} \hline
{\bf Species} &{\bf Observations} & {\bf This Work}& {\bf Time of best fit} &{\bf Ref}\\
&& {\bf M1}& {\bf yrs} &\\ \hline
{\bf DCO$^{+}$}& 7.5(-12)& $\sim$1 - 5(-11)& $\le$9.00(3) & \citet{vand95}\\ 
{\bf DCN}&1.3(-11)& 1.3 - 5.7(-10)& 1.7 - 9(4)& \citet{vand95}\\ 
{\bf DNC}&2.5(-12)& $\le$1(-13)& all times & \citet{vand95} \\ 
{\bf HDS}&7.5(-11)& 1(-7) - 1(-10)& 6(4) - 2.5(5)& \citet{vand95}  \\
{\bf HDO} &9(-8) &$\le$1(-7)& all times &\citet{cout13}\\
{\bf NH$_2$D }&5(-10)& 4(-8) - 1(-10)& 9(4) - 9(5)& \citet{vand95}\\
{\bf HDCO}&6(-11)&$\le$6(-11)& 1.8(4) \& $\ge$2.5(5) & \citet{vand95}\\
          &      &$\ge$6(-11)& 0.17 - 2.5(5) & \\
{\bf $^{\dag}$D$_2$CO} &5(-10) & $\sim$4(-10)& 0.2 - 2.4(5)& \citet{cec98c} \\
{\bf$^{\dag}$CH$_3$OD}&3.6(-10)& $\le$4(-11)& 0.9 - 5.7(5)&\citet{par02}\\
{\bf $^{\dag}$CH$_2$DOH}&7.5(-9)&$\le$9(-11)&1.4 - 3.5(5)& \citet{par02}\\
{\bf C$_2$D }&2.3(-11) & 1 - 4.5(-11)&3.8 - 5.5(4)&\citet{vand95}\\ 
{\bf $^{\dag}$DCOOCH$_3$}&1.5(-9) &$\le$1(-13) & all times&\citet{dem10}\\
 \hline \hline
{\bf Ratio}         & & & & \\ \hline
{\bf DCO$^{+}$/HCO$^{+}$}  &0.009           & 0.009        & $\sim$ 4.03(4)    &\citet{vand95}\\
{\bf DCN/HCN}            & 0.013          & 0.012         & 2.6(4)            &\citet{vand95}\\
{\bf DNC/HNC}            & 0.03           & $\le$1(-7)    & all times         & \citet{vand95}\\
{\bf C$_2$D/C$_2$H}      & 0.18            & 0.18          & 4.7(4) 	      & \citet{vand95}\\
{\bf HDCO/H$_2$CO}       & $\sim$0.14      & few(-1)       & 2(3) - 1.6(4)    & \citet{rou20}\\
{\bf D$_2$CO/H$_2$CO}    & $\le$0.1        & $\le$0.1      & $\ge$1.7(4)      & \citet{cec98c}\\
{\bf D$_2$CO/HDCO}       & $\le$0.5        & $\le$0.5      & 1.8(4) - 1.32(5) & \citet{rou20}\\
{\bf NH$_2$D/NH$_3$}     & $\sim$0.1       & $\le$1(-4)    & all times 	      & \citet{rou20} \\
{\bf HDO/H$_2$O}         & 0.04 - 0.51   & 1 - 2         & all times        & \citet{cout13} \\
{\bf CH$_3$OD/CH$_3$OH}  & 0.02      & $\le$0.1      & $\ge$ 1.74(4)    & \citet{par04}\\
{\bf CH$_2$DOH/CH$_3$OH} & 0.30           & 1-9(-1)       & $\ge$ 1.74(4)    & \citet{par04}\\
{\bf HDS/H$_2$S}         & 0.1             & 0.15$\pm$0.01 & 6.9 - 9.1(4)     & \citet{vand95} \\ 
{\bf $^{\ddag}$N$_2$D$^{+}$/N$_2$H$^{+}$} & 0.25 & $\ll$0.25 & all times & \citet{rob07} \\ \hline \hline
\end{tabular} 
\flushleft
$^{\dag}$ This value is converted into fractional abundance assuming N(H$_2$) = 2$\times$10$^{23}$ cm$^{-2}$ as given by \citet{vand95}.\\
a(b) means a$\times$10$^b$\\
$^{\ddag}$ Ratio observed in NGC 1333 IRAS4 by \citet{rob07}
\end{table*}

Most of the species in our sample match observations at times between 10$^{4}$ and 
10$^{5}$ years, which is the assumed age for a typical Class 0 source \citep{and93}. 
Exceptions are the ions that fit observations at times earlier than 10$^{4}$ years 
and show a rapid decrease after that time, and HDO that does not match observations 
at any time and is always underestimated by our model.

Deuterated ammonia matches observations at times later than 9$\times$10$^{4}$ years. 
Currently, our model is underestimating the abundances of some deuterated species by one or 
two orders of magnitude, such as CH$_3$OD, CH$_2$DOH, DNC, and DCOOCH$_3$. 
The main reason for this is perhaps the exclusion of multiply deuterated H$_3^+$ isotopologues 
that play a role in enhancing the deuterium fractionation via gas-phase chemistry \citep{rob03}. 
 
The N$_2$D$^{+}$/N$_2$H$^{+}$ ratio matches observations only during early times
$\sim$4$\times$10$^3$ years. In fact, \citet{rob07} surveyed the N$_2$D$^{+}$/N$_2$H$^{+}$ ratio around a
sample of Class 0 sources and concluded that the high ratio observed represents the cooler gas in the
extended envelope around the source and not the hot gas in the core. Similar results were reported
by \citet{emp09} in their survey for N$_2$D$^+$/N$_2$H$^+$ in a sample of Class 0 sources. 
Yet our calculated HDO/H$_2$O ratio does not match observations. This must be because 
our model underestimates the fractionation by at least a factor of two which could be a result of our model limitations, 
described earlier in \S \ref{model}.

{\it \bf The Orion Kleinmann-Low nebula (Orion KL)} is the nearest massive star-forming region at $\sim$ 410 pc. The core 
has physical conditions (n(H$_2$) = 10$^7$ {\bf cm$^{-3}$}, T $\sim$ 150-300K, mass $\sim$ 15 M$_{\odot}$; 
see \citealt{neil13a, kau98} and references therein) close to those of our model M7. 
Most recently, \citet{neil13a, neil13b} surveyed deuterated molecules in the Orion KL region using the Herschel/HIFI facilities. 
Table (\ref{tab:Dratio}) summarizes a comparison between fractional abundances {\bf and molecular ratios} observed in the Orion 
hot core region and our model, M7, calculations. The table also lists times of best fit. 
Apart from CH$_2$DOH our model seems to be consistent with the observations reported by \citet{neil13a, neil13b}.
\begin{table*}
\centering
\caption{The calculated deuterium fractional abundances and molecular D/H ratios in hot core 
Model M7 (Phase II) in comparison with observations of the Orion KL hot core.} 
\label{tab:Dratio}
\begin{tabular}{lllll} \hline
{\bf Species} &{\bf Observations} & {\bf This Work}& {\bf Time of best fit} &{\bf Ref}\\
&& {\bf M7}& {\bf yrs} &\\ \hline
{\bf HDO} & $\sim$ 4.5 (-8) &$\ge$4.2(-8) & $\ge$3.8(4)&\citet{neil13a}\\
{\bf NH$_2$D }& 1.4(-8)& $\ge$1(-8)& 0.3 - 5.8(5) & \citet{neil13b}\\
{\bf HDCO}&2.9 (-10)&$\sim$1(-10)-3(-8)& $\ge$1.3(4)& \citet{neil13b}\\
{\bf CH$_3$OD}& 1.9(-9)& $\le$1(-10) & all times &\citet{neil13b}\\
{\bf CH$_2$DOH}& 4.5(-9)& $\le$1(-10) & all times & \citet{neil13b}\\
\hline \hline
{\bf Ratio}         & & & &\\ \hline
{\bf HDO/H$_2$O}     & $\sim$3$^{+3.1}_{-1.7}$(-3)& 1.3-6(-3) & 1.6-8.9(5) & \citet{neil13a} \\
{\bf NH$_2$D/NH$_3$} & $\sim$6.8$\pm$2.4(-3) & 4.5-9.2(-3) & 1.5-1.9(4)& \citet{neil13b} \\
{\bf HDCO/H$_2$CO}  & $\le$0.005      & $\sim$5(-3) &1.05-1.15(5) & \citet{neil13b}\\
{\bf CH$_3$OD/CH$_3$OH}& $\le$1.8(-3) & $\le$1.8(-3) & $\ge$4.92(5) & \citet{neil13b}\\
{\bf CH$_2$DOH/CH$_3$OH} & $\le$4.2(-3) & $\le$4.2(-3)& $\ge$3.29(5) & \citet{neil13b}\\
\hline\hline
\end{tabular} 
\flushleft
a(b) means a$\times$10$^b$\\ 
\end{table*}
\subsection{Comparison with previous astrochemical modelling}
\label{comp-mod}
In this section, we also briefly compare our model results with other astrochemical modelling efforts. 
\citet{albrt013} 
performed a detailed chemical study of deuterated molecules using gas-grain chemical models. 
They estimate the abundances of HDO in hot cores and corinos to be $\sim 10^{-7}$. 
This value best fits our models for times around 1-5$\times10^{5}$ yrs for hot corinos (e.g. Fig. 1-a) 
and time ranges of 10$^{5.2}$-10$^{6}$ yrs, and  10$^{4.3}$-10$^{6}$ yrs for hot cores of 5 and 25M$_{\odot}$, respectively.

\citet{aik012} 
used a gas-grain chemical model of an in-falling parcel of fluid to study the evolution of deuterated species 
and the deuterium fragmentation from pre- to proto-stellar cores. They found that as the gas depletion increases, 
the deuterium fraction enhances. 
Our calculations for both hot corinos, Model M3, and hot cores, Models M6 and M9, showed that for most of the 
studied species in both environments, the fractional abundances are enhanced as the gas depletion percentage increases; 
see Figs. (\ref{fig:3} and \ref{fig:4}). These results are supported by the model calculations of \citet{aik012}.

\section{Conclusion}
\label{conc}
We modelled the deuterium chemistry in hot corinos and cores using an extended, updated and improved version 
of the chemical model used by \citet{viti04} for hot cores and modified for hot corinos as described in \citet{awad10}. 
Unlike previous studies, here we focus on the protostellar phase 
where the evaporation of mantle species occurs and influences the chemistry of the core. 
The novelty of our approach is the treatment of the evaporation of deuterated species 
adopting the experimental results of \citet{col03a, col04}. We studied the 
influence of varying the physical conditions of star forming regions, namely the density and 
depletion efficiency, on the chemical evolution of the core. In addition, we explored the 
effect of changing the collapse timescale of low-mass stellar cores, as well as the final mass of the protostar. 

For both hot core and corino environments, we found that lowering the depletion 
percentage decreases the abundance of most of the studied deuterated species, with 
the exception of HDCS in both cores and D$_2$CO in hot cores. 
Deuterated species in hot cores are more 
sensitive to the changes in the cores density than in hot corinos. Generally, we 
find that decreasing the density of the gas reduces the abundances of the deuterated species, 
in particular large species, in hot cores more than in hot corinos. 

In addition, in hot corinos, our models showed that the collapse time affects the abundance of HDCS, NH$_{2}$D, CH$_2$DOH, and HDO, 
so that this should also be taken into account when attempting to model the deuteration of water, besides taking into account 
variations in physical parameters and dust temperature as already explored by \citet{taq12aa} and \citet{caz11apj}.

Our model failed to reproduce the observed abundances of large organic species such as HCOOCH$_3$, and its deuterated 
counterpart. This is most likely due to an incomplete surface 
chemistry as well as the lack of multiply deuterated species in our chemical network. 
\section*{Acknowledgments}
Z.Awad is grateful to Professor C. Ceccarelli for her helpful suggestions to improve 
the results and discussion part. E. Bayet thanks STFC astrophysics at Oxford 2010-2015 
(ref: ST/H002456/1) and John Fill OUP research fund "Molecules in galaxies: securing Oxford's 
position in the ALMA era" (ref: 0921267). The research leading to these results has received 
funding from the (European Community’s) Seventh Framework Program [FP7/2007–2013] under grant 
agreement no. 238258. P. Caselli acknowledges the financial support of the European Research 
Council (ERC; project PALs 320620). S. Viti and P. Caselli acknowledge support the UK Science 
and Technology Funding Council.

\newcommand{\apj}[1]{ApJ, }
\newcommand{\jcp}[1]{J. Chem. Phys., }
\newcommand{\mnras}[1]{MNRAS, }
\newcommand{\aj}[1]{Aj, }
\newcommand{\apjs}[1]{ApJS, }
\newcommand{\apjl}[1]{ApJ Letter, }
\newcommand{\aap}[1]{A\&A, }
\newcommand{\aaps}[1]{A\&A Suppl. Series, }
\newcommand{\araa}[1]{Annu. Rev. A\&A, }
\newcommand{\aaas}[1]{A\&AS, }
\newcommand{\apss}[1]{Ap\&SS }
\newcommand{\bain}[1]{Bul. of the Astron. Inst. of the Netherlands,}
\newcommand{\planss}[1]{Planetary and Space Science,}
\newcommand{\nat}[1]{Nature,}
\newcommand{\aapr}[1]{The Astronomy and Astrophysics Review,}
\bibliographystyle{mn2e} 
\bibliography{D-references}

\end{document}